\documentclass[12pt,english,aps,manuscript]{revtex4}
\usepackage[T1]{fontenc}
\usepackage[latin9]{inputenc}
\usepackage[a4paper]{geometry}
\usepackage{babel}
\usepackage{float}
\usepackage{amsmath}
\usepackage{graphicx}
\usepackage[unicode=true,
 bookmarks=true,bookmarksnumbered=true,bookmarksopen=false,
 breaklinks=false,pdfborder={0 0 0},backref=false,colorlinks=false]
 {hyperref}
\hypersetup{
 pdfauthor={Jakub Krajniak},
 pdfkeywords={praca magisterska}}

\makeatletter
\@ifundefined{textcolor}{}
{%
 \definecolor{BLACK}{gray}{0}
 \definecolor{WHITE}{gray}{1}
 \definecolor{RED}{rgb}{1,0,0}
 \definecolor{GREEN}{rgb}{0,1,0}
 \definecolor{BLUE}{rgb}{0,0,1}
 \definecolor{CYAN}{cmyk}{1,0,0,0}
 \definecolor{MAGENTA}{cmyk}{0,1,0,0}
 \definecolor{YELLOW}{cmyk}{0,0,1,0}
}
\numberwithin{equation}{section}

\@ifundefined{date}{}{\date{}}
\AtBeginDocument{
  
}

\makeatother

\begin{document}

\title{Monte Carlo Study of Patchy Nanostructures Self-Assembled from a
Single Multiblock Chain}

\author{J. Krajniak, M. Banaszak{*}}

\address{Faculty of Physics, A. Mickiewicz University ul. Umultowska 85, 61-614
Poznan, Poland}

\email{mbanasz@amu.edu.pl}

\selectlanguage{english}%
\begin{abstract}
We present a lattice Monte Carlo simulation for a multiblock copolymer
chain of length N=240 and microarchitecture $(10-10)_{12}$. The simulation
was performed using the Monte Carlo method with the Metropolis algorithm.
We measured average energy, heat capacity, the mean squared radius
of gyration, and the histogram of cluster count distribution. Those
quantities were investigated as a function of temperature and incompatibility
between segments, quantified by parameter $\omega$. We determined
the temperature of the coil-globule transition and constructed the
phase diagram exhibiting a variety of patchy nanostructures. The presented
results yield a qualitative agreement with those of the off-lattice
Monte Carlo method reported earlier, with a significant exception
for small incompatibilities, $\omega$, and low temperatures, where
3-cluster patchy nanostructures are observed in contrast to the 2-cluster
structures observed for the off-lattice $(10-10)_{12}$ chain. We
attribute this difference to a considerable stiffness of lattice chains
in comparison to that of the off-lattice chains.
\end{abstract}

\date{July 29, 2013}

\maketitle

\section{INTRODUCTION}

Block copolymers are studied mainly due to a fundamental interest
in the soft matter research and also in the industrial applications
\citet{2004}. he physical properties, such as elasticity, toughness
or electrical conductivity, depend mainly on chemical composition
and microarchitecture of the polymer chain. Numerous studies show
that diblock polymer melts can spontaneously form variety of nanostructures
such as: ordered layers, hexagonally packed cylinders, cubically ordered
spheres and the celebrated gyroid structures \citep{Leibler1980,Matsen1994}.
It has been shown that one can also obtain different phases by varying
the chain microarchitecture\citep{2004,Binder2008,PhysRevE.66.031804,Banaszak2003}.
We expect that increasing the complexity of microarchitecture leads
to more nanostructures. Indeed, triblock copolymers show new phases
like lamella-cylinder or lamella-sphere combinations, which have been
confirmed experimentally. Synthesizing more complex microarchitectures,
including the cyclic and the branched ones, results in a plethora
of new phases \citep{Knoll2004}. Moreover, not only polymer melts
are prone to self-assembly into various phases but also a single copolymer
polymer chain can self-assemble into various structures, referred
to as patchy nanostructures or patchy particles \citep{Zhang2004}.
Recent studies in both computer simu- lation and supramolecular chemistry
show that such systems are thermodynamically stable and, more importantly,
can be obtained by chemical synthesis \citep{Zhang2004,Rupar2012}.
Such patchy nanostructures can be used as building blocks for self-assembling
nanodevices. It has been shown that due to well defined symmetry of
such particles they can form aggregates with an ordered spatial structure.
This behavior can be used for the bottom-up approach, overcoming the
size limits of producing particles of sizes from 7 to 17 nm with a
defined symmetry, in one step \citep{Rupar2012,Zhang2011}. In previous
studies, using an off-lattice Monte Carlo method with a discontinuous
(square- well) potential \citep{PhysRevE.84.011806} and using the
Lennard-Jones (LJ) potential \citep{PhysRevLett.99.228302}, the stability
of patchy nanostructures was investigated. In order to probe the free
energy landscape more efficiently, the parallel tempering (PT) method
\citep{Katzgraber2006,Sikorski2002,Gront2007,Lewandowski2010,Beardsley2010}
was employed in ref. \citep{PhysRevE.84.011806} and Wang-Landau method
\citep{PhysRevLett.86.2050} in ref. \citep{PhysRevLett.99.228302}.
In this study we intend to study one of the chain microarchitectures,
specifically $(10-10)_{12}$ , but using a lattice Monte Carlo method.
While the off-lattice models are more realistic, they also require
more computational effort. Therefore it is significant to asses the
relative merits of the lattice model. If the lattice model gives similar
results, then it may be reasonable to use the lattice model rather
than the off-lattice models. The aim of this paper is to answer the
following questions: 
\begin{itemize}
\item are the nanostructures obtained in the lattice simulation the same
as those obtained in the off-lattice simulation?
\item is the phase diagram obtained in the lattice simulation the same as
that obtained in the off-lattice simulation?
\end{itemize}

\section{Model and method}

In this study, a coarse-grained model is used. The polymer chain is
placed on the face centered cubic (FCC) lattice with coordination
number z=12 and the bond length equal to $\sqrt{2}a$. Chain bonds
are not allowed to be broken or stretched. The periodic boundary conditions
are applied. The size of simulation box is chosen to fit a fully extended
chain. Polymer-solvent interactions are included in an implicit manner
in the polymer-polymer interaction potential. Polymer chain consists
of two types of monomers: A and B. Interaction energy between monomers
is defined as follows: $\epsilon_{AA}=\epsilon_{BB}=-\epsilon$ and
$\epsilon_{AB}\in\langle-\epsilon,-0.1\epsilon\rangle$. The $\epsilon$
parameter is positive and serves here as an energy unit. We define
reduced energy as 

\begin{equation}
E^{*}/N=E/(\epsilon N)
\end{equation}

and reduced temperature as

\begin{equation}
T^{*}=k_{B}T/\epsilon
\end{equation}

where N is the number of chain monomers and $k_{B}$ is the Boltzmann
constant. Negative value of interaction energy means that there is
a net attraction between monomers. Reduced temperature parameter is
used to control the quality of solvent, from good to bad which causes
a transition from swollen state to a globular state. Dimensionless
parameter $\omega=-\epsilon_{AB}/\epsilon$ is a measure of compatibility
between two monomers of different type. Lower values of $\omega$
mean that an attraction between monomers A and B is lower than between
monomers of the same type. For $\omega=1$ the copolymer chain becomes
a homopolymer chain. The chain consists of 120 monomers of type A
and 120 monomers of type B which form the multiblock microarchitecture
$(10-10)_{12}$. During the simulation various $T^{*}$\textquoteright s
and $\omega$\textquoteright s are probed; $T^{*}$ from 2.27 to 20.0
(because below 2.27 nothing seems to change in the polymer structure),
and $\omega$ from 0.1 to 1.0.

The simulation was performed by standard Monte Carlo simulation method
with Metropolis acceptance criteria \citep{Metropolis1953}. In order
to perform a Monte Carlo move we use pull-move algorithm, which employs
chain movements, reminiscent of the reptation moves, by pulling the
chain in a random direction as described in detail in reference \citep{Lesh03acomplete}.
We define one Monte Carlo Step (MCS) as the attempt to perform one
move according to the algorithm. Each simulation consists of $4\times10^{7}$
MCS. First the system is equilibrated athermally, and next $2\times10^{7}$
steps are run in the thermal condition. We calculate the thermal averages
from last $2\times10^{7}$ steps.

\section{Results and discussion}

\begin{figure}[H]
\includegraphics[width=0.5\textwidth]{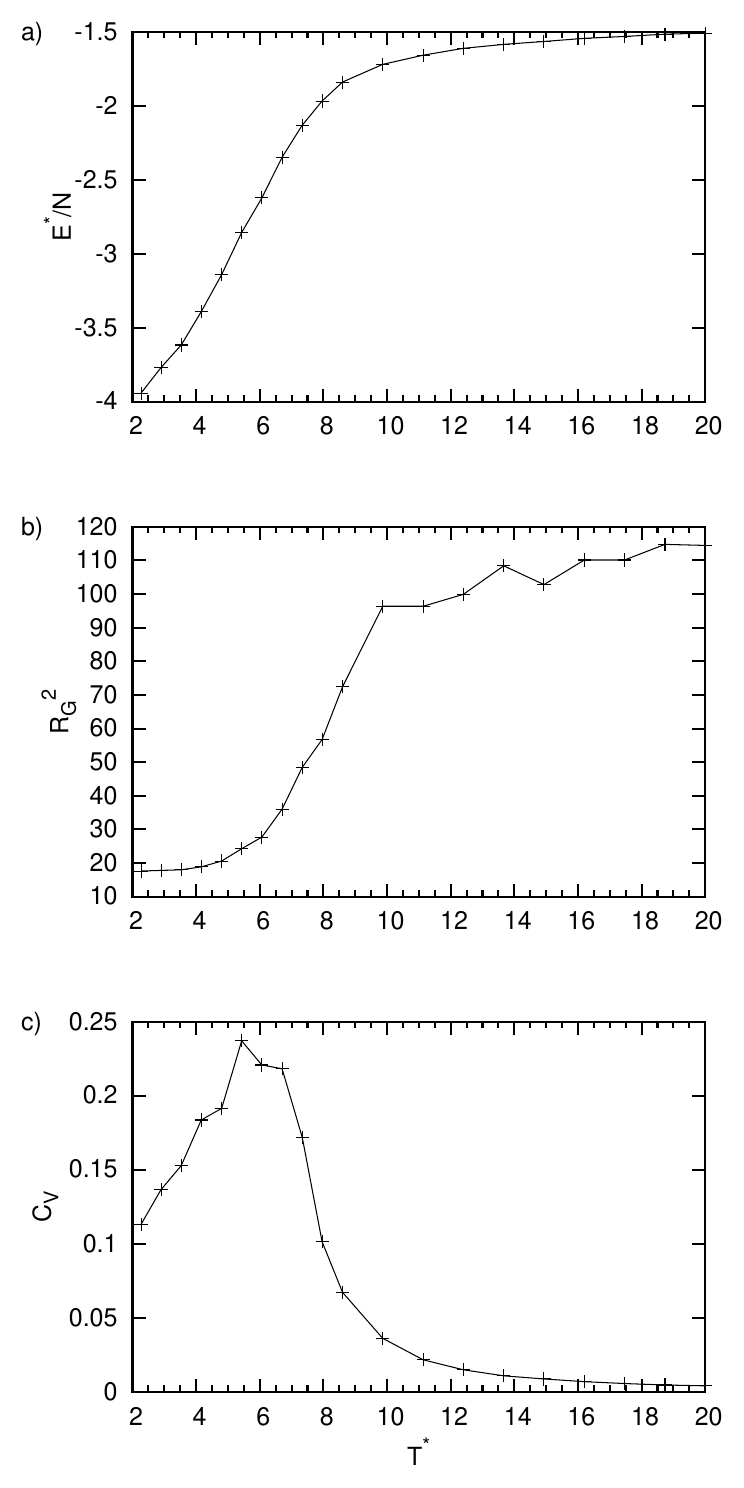}

\protect\caption{Reduced energy per monomer $E^{*}/N$ (a), squared radius of gyration
$R_{G}$ (b) and specific heat $C_{V}$ (c) as a function of T for
$\omega=0.1$.\label{fig:fig1}}
\end{figure}

We start the simulation by equilibrating the system in the athermal
limit, where $\epsilon/k_{B}T=0$. After the system reaches its thermal
equilibrium we start to cool it down to a variety of $T^{*}$\textquoteright s.
First, we present results for the lowest compatibility parameter $\omega=0.1$.
As $T^{*}$ is decreased, it is expected that the chain undergoes
a transition from a swollen state to an intermediate state pearl-necklace,
and after that to the globular state \citep{Lewandowski2008}. In
figure (\ref{fig:fig1}) we show the Monte Carlo results for the energy
per monomer, $E^{*}/N$, mean squared radius of gyration, $R_{G}^{2}$,
and heat capacity, $C_{V}$, of the polymer chain, as a function of
$T^{*}$. The heat capacity, $C_{V}$, is obtained from the energy
fluctuations as follows 
\begin{equation}
C_{V}=\frac{\langle(E^{*}-\langle E\rangle)^{2}\rangle}{NT^{*2}}
\end{equation}

where N is the number of monomers, and $\langle\ldots\rangle$ denotes
the thermal average. The reduced energy does not change much from
high temperatures to about $T^{*}=10.0$ where it starts to decrease.
A similar behavior is observed for mean squared radius of gyration
which also decreases below $T^{*}=10.0$. In figure (\ref{fig:fig1})c)
we show the temperature dependence for the heat capacity, $C_{V}$
. The maximum in $C_{V}$ corresponds approximately to the inflection
point in the reduced energy presented in figure (\ref{fig:fig1})a).
From those two observations we can estimate the coil-to-globule transition
temperature, $T_{CG}^{*}=5.43$. In the figure (\ref{fig:fig2}) we
present the equilibrated structures in a coiled state (a), in a state
which is close coil-to-globule transition (b) and in a globular state
(c) for $\omega=0.1$. As expected, upon cooling the chain collapses.
\begin{figure}[H]
\includegraphics[width=0.5\textwidth]{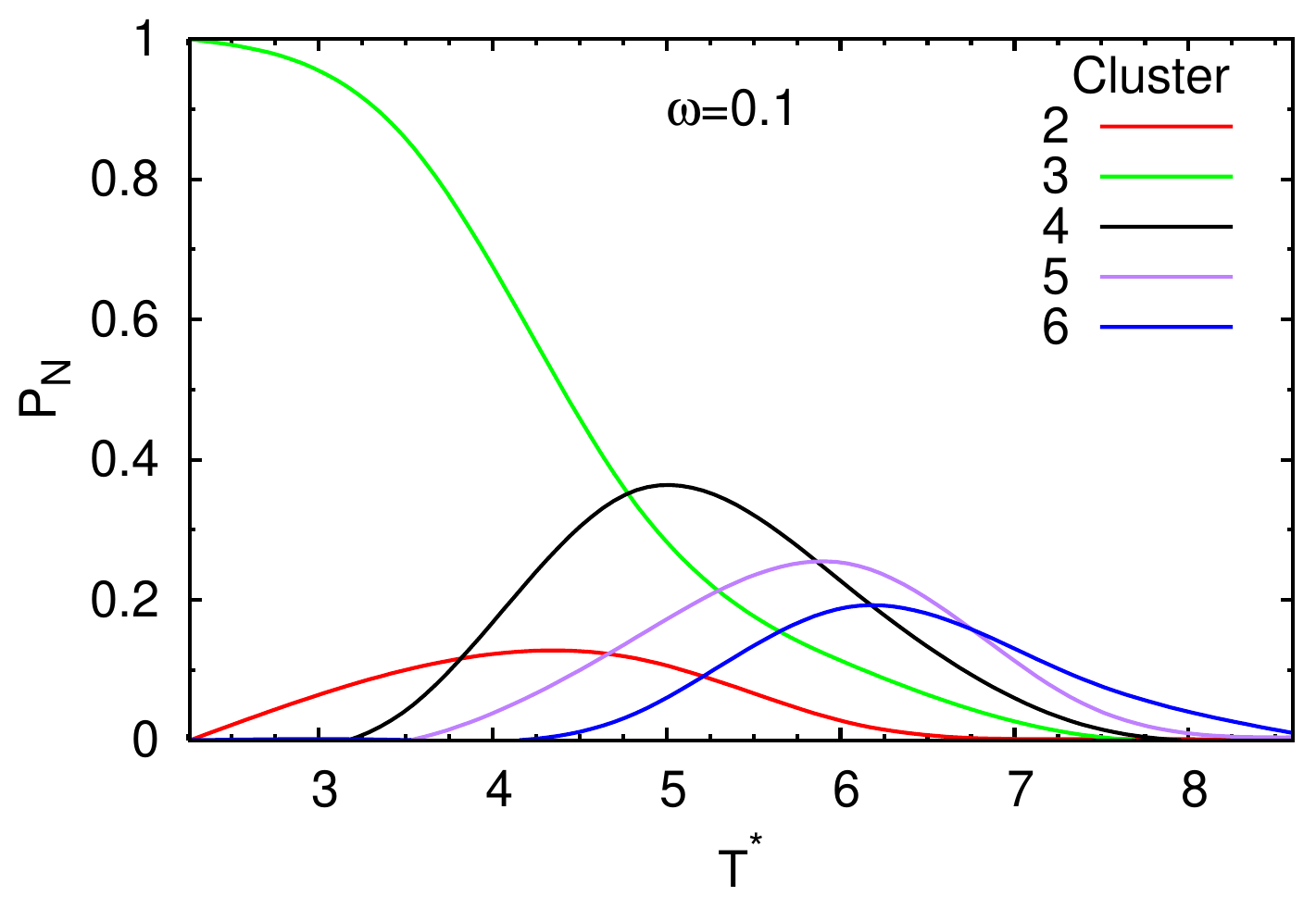}

\protect\caption{Probability of finding structures with n-clusters for $\omega=0.1$.\label{fig:fig3}}
\end{figure}

In order to better distinguish different patchy nanostructures, we
use cluster count distribution as in ref. \citep{PhysRevE.84.011806}.
In figure (\ref{fig:fig3}), we present histogram of probability of
clusters with number of segments set to 2, 3, 4, 5, and 6, for $\omega=0.1$.
We show that the 6-cluster structure (and also n-cluster structures,
with $n>6$) is most stable at $T^{*}$\textquoteright s that higher
than $T^{*}=6.75$, but from $T^{*}=5.90$ to $T^{*}=6.75$, the 5-clusters
occur with the highest probability. Next from $T^{*}=4.75$ to $T^{*}=5.90$
the 4-clusters prevail. It is also worthwhile to notice that within
this range falls the coil-to-globule transition, $T^{*}=5.43$. For
$T^{*}<4.75$ the 3-clusters are the most probable patchy nanostructures.
The representative 2-clusters are shown in figure (\ref{fig:fig4})
and they seem to be similar to the lamellar nanophase which is observed
in diblock copolymer melts. Probability of observing the 3-clusters
gradually increases, reaching unity at low temperatures. Variations
in number of clusters and, as a result, in the patchy nanostructure
can also be discerned by measuring the heat capacity with higher temperature
resolution (more $T^{*}$\textquoteright s), but additional Monte
Carlo simulations would be required.
\begin{figure}[H]
\begin{centering}
\includegraphics{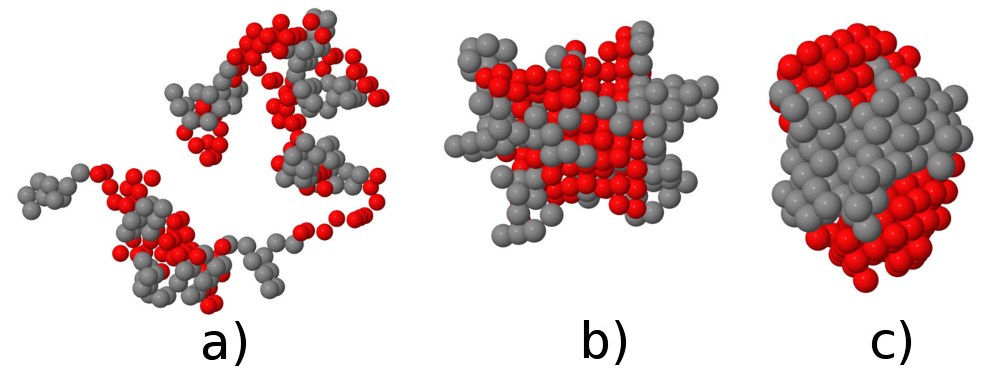}
\par\end{centering}

\protect\caption{Sample snapshots of the chain for $\omega=0.1$ at different temperatures:
swollen state at $T^{*}=13.67$ (a), coil-to-globule transition at
$T^{*}=5.43$ (b) and 3-cluster structure at $T^{*}=3.53$ (c).\label{fig:fig2}}
\end{figure}
\begin{figure}[H]
\begin{centering}
\includegraphics{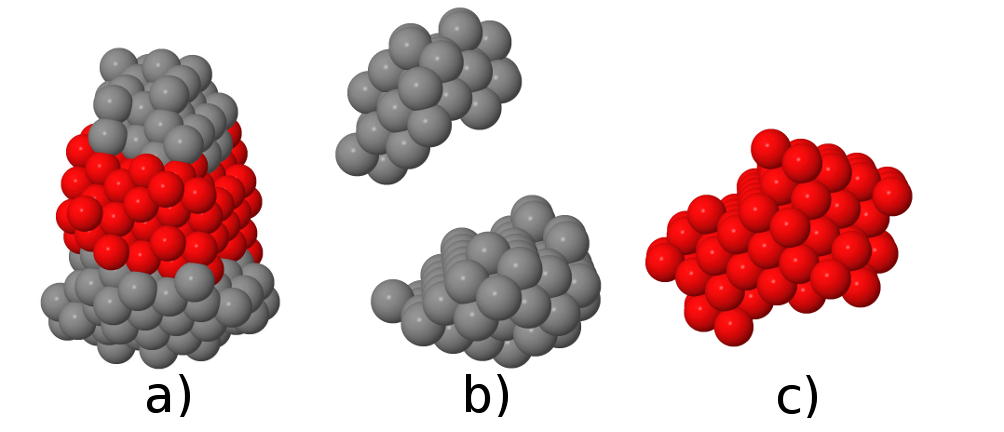}
\par\end{centering}

\protect\caption{Example of 3-cluster structure for $\omega=0.1$ and $T^{*}=2.27$:
(a) - whole molecule, (b) - structure with hidden segments of type
A, (c) - molecule with hidden segments of type B.\label{fig:fig4}}
\end{figure}

\begin{figure}[H]
\includegraphics[width=0.5\textwidth]{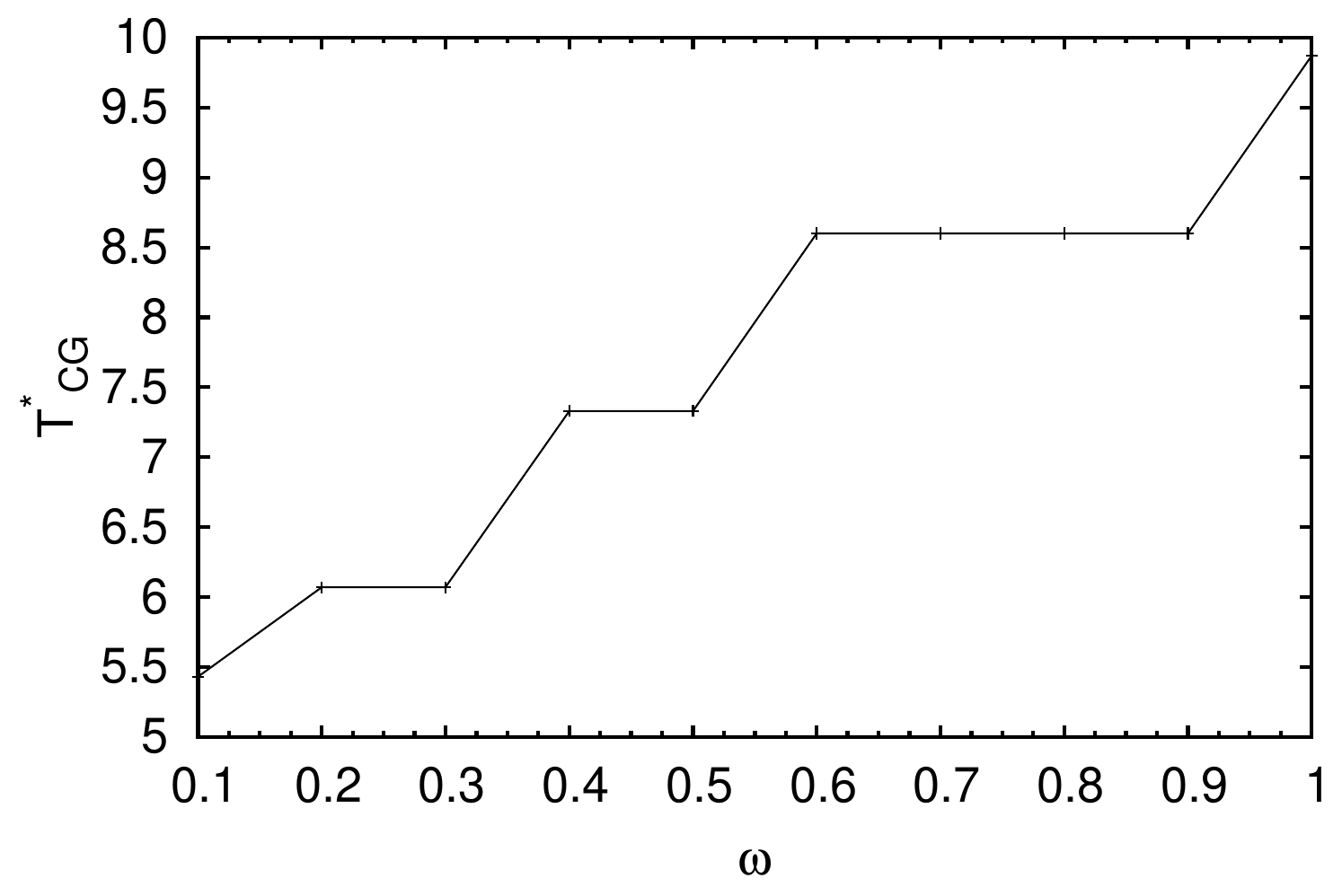}

\protect\caption{The reduced temperature $T_{CG}^{*}$ as the function of $\omega$.\label{fig:fig5}}
\end{figure}

Next, we describe the simulation results for $\omega$\textquoteright s
from 0.2 to 1.0, in more detail. In figure (\ref{fig:fig5}) we show
the $\omega$ dependence of the coil-to-globule transition temperature,
$T_{CG}^{*}$, which is similar to that observed in ref. \citep{PhysRevE.84.011806},
indicating that $T_{CG}^{*}$ increases upon increasing $\omega$.

\begin{figure}[H]
\includegraphics[width=0.5\textwidth]{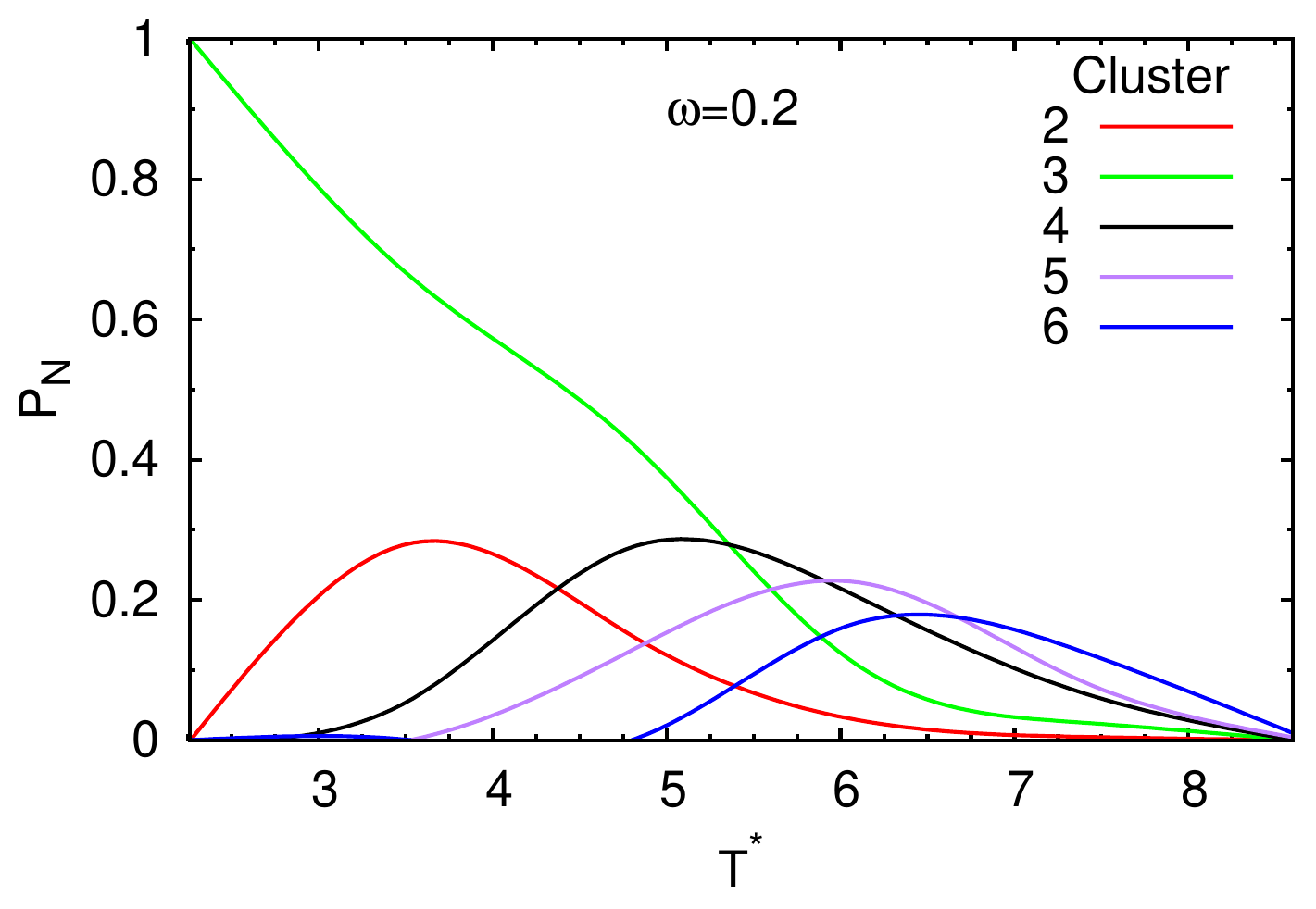}

\protect\caption{Probability of finding n-cluster structure for $\omega=0.2$.\label{fig:fig6}}
\end{figure}

Similarly as for $\omega=0.1$, we determine the temperature dependencies
for the heat capacity, the mean squared radius of gyration and the
energy per monomer. In figure (\ref{fig:fig6}) we present cluster
count distribution for $\omega=0.2$. In this case we also observe
that the 3-cluster structure has the highest probability at low $T^{*}$
but the probability of obtaining the 2-cluster structures increases
to 0.3 at $T^{*}=3.51$. Transition between 4- and 5-cluster is shifted
towards higher temperatures, as expected. The $T^{*}$ range for 4-clusters
shrinks significantly from (4.75, 5.90) for $\omega=0.1$ to (5.5,
5.92) for $\omega=0.2$ . The transition temperature from 4-cluster
to 5-cluster patchy structure ($T=5.92$) is close to the coil-to-globule
transition temperature, $T^{*}=6.06$. Above that temperature there
is no obvious prevalence of any structure as it becomes increasingly
disordered upon heating.

\begin{figure}[H]
\includegraphics[width=0.5\textwidth]{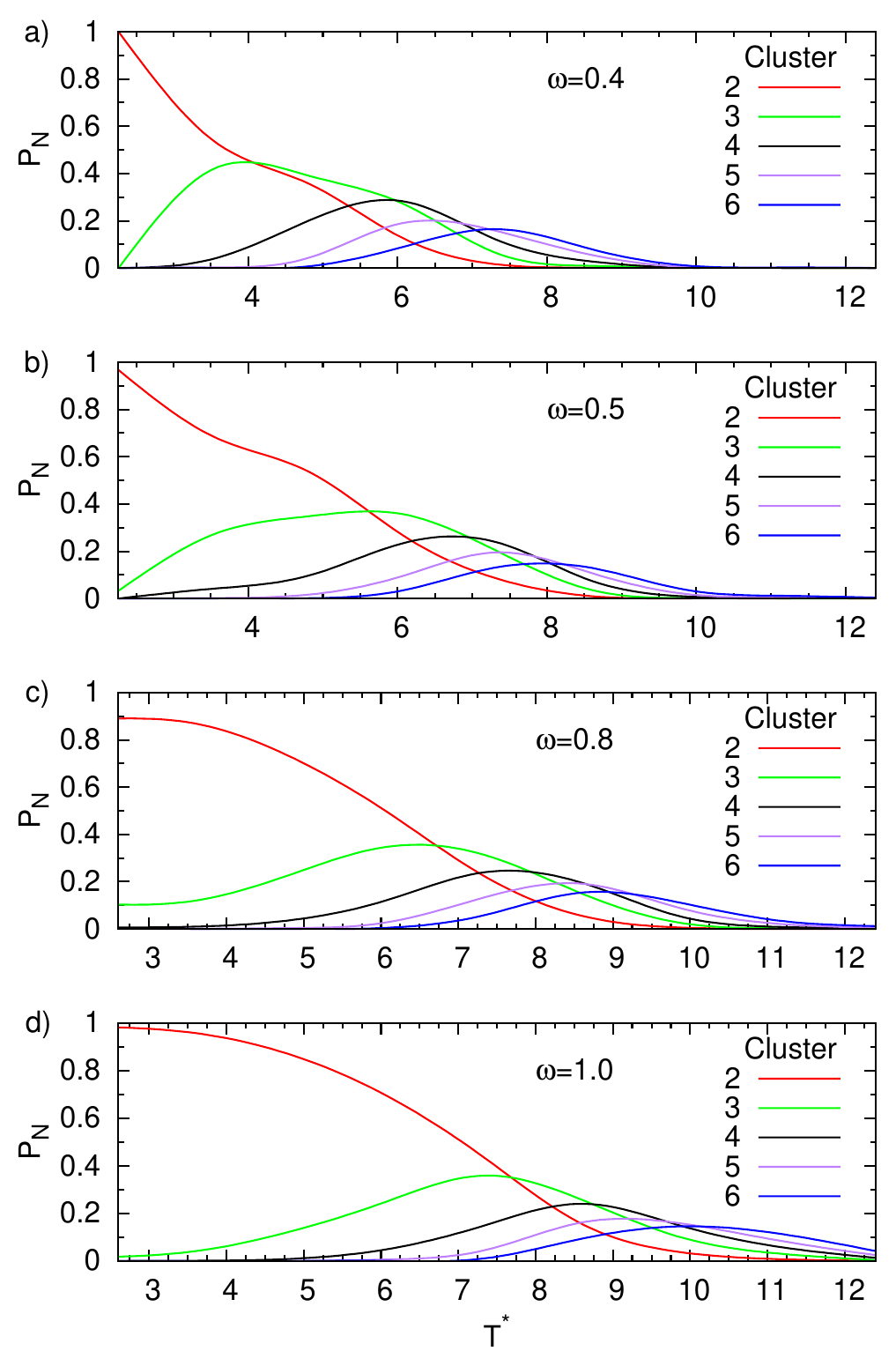}

\protect\caption{Probability of finding different cluster structures as a function
of reduced temperature T {*} with compatibility: (a) $\omega=0.4$,
(b) $\omega=0.5$, (c) $\omega=0.8$, (d) $\omega=1.0$.\label{fig:fig7}}
\end{figure}

\begin{figure}[H]
\includegraphics[width=0.5\textwidth]{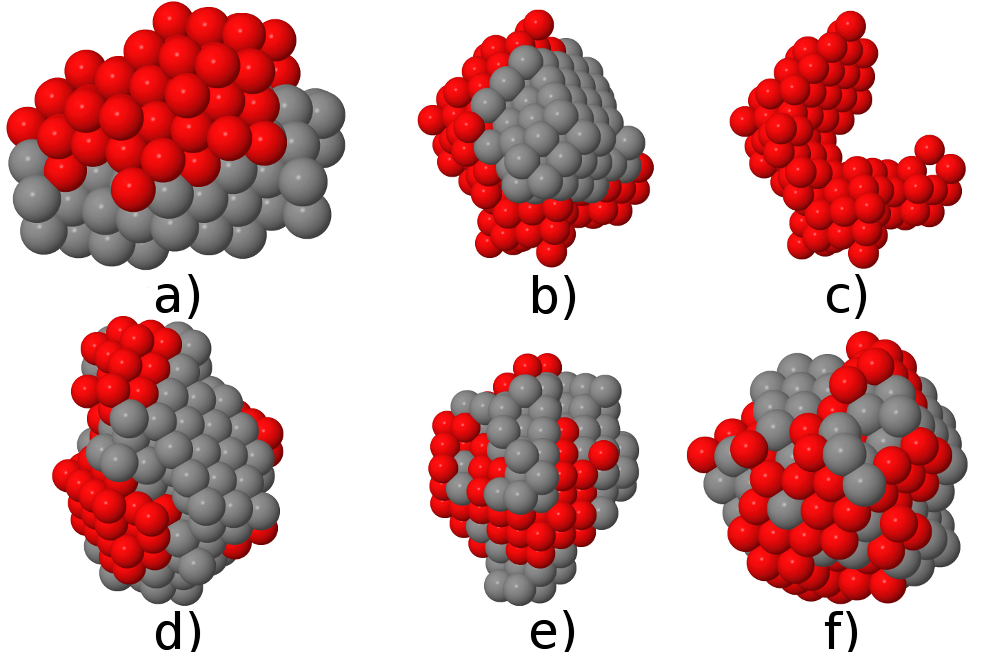}

\protect\caption{Variety of 2-cluster structures for the $(10-10)_{12}$ chain at $T^{*}=2.27$
for compatibility parameter: (a) $\omega=0.4$ double droplet, (b)
$\omega=0.5$ hand-shake (c) $\omega=0.5$ (without the B segments)
(d) $\omega=0.7$, (e) $\omega=0.9$, (f) $\omega=1.0$.\label{fig:fig8}}
\end{figure}

In figure (\ref{fig:fig7}) we show cluster count distribution for
the representative $\omega$'s: 0.4, 0.5, 0.8 and 1.0. We observe
that the probability maxima for different cluster counts (for $n>3$)
are shifted towards higher temperatures as $\omega$ is increased.
It is interesting to identify the most probable patchy nanostructures
at low temperatures, as they may correspond to the native states in
biopolymers \citep{Shakhnovich2006}. The most probable structures
for $\omega=$0.1, 0.2 and 0.3 are the 3-clusters. On the other hand,
for $\omega$\textquoteright s which are equal or greater than 0.4
the 2-cluster are most probable. However, it is interesting to record
that the probability of the dominant nanostructures, either 2-cluster
or 3-cluster, approaches unity at low temperatures. Sample snapshots
of molecules for various $\omega$\textquoteright s at $T^{*}=2.27$
are shown in figure (\ref{fig:fig8})a). For lower $\omega$\textquoteright s
segments are strongly segregated which leads to formation of a nanostructure
consisting of two hemispheres (double-drop structure). Other structures
which can be observed are the hand-shake-like structures which are
shown in figure (\ref{fig:fig8})b and 8c. Above $\omega=0.7$ the
structures become more disordered.

In this paper as well as in the previous study \citep{PhysRevE.84.011806}
we do not observe a single native state for the polymer chain. Structures
with different number of clusters can coexists. Only for the lowest
temperature $T^{*}$ the probability for a given number of clusters
is approximately 1.0. Therefore we can conjecture that instead of
one native state for given ($\omega,\:T^{*}$) we obtain numerous
stable states with different spatial order but with the same energy.
In order to compare the present lattice model with the off-lattice
model we construct a phase diagram in the ($\omega,\:T^{*}$) space.
According to previous multiblock studies we expect that those two
models should give similar results \citep{Lewandowski2008}. In particular,
from ref. \citep{Lewandowski2008} we learn that in order to match
the coil-to-globule transision temperature for lattice model, $T_{CG}^{*latt}$
, with that for the off-lattice model, $T_{CG}^{*off-latt}$ , we
need to use a multiplicative factor, R, defined below:
\begin{equation}
T_{CG}^{*latt}=R\times T_{CG}^{*off-latt}
\end{equation}

\begin{figure}[H]
\includegraphics[width=0.5\textwidth]{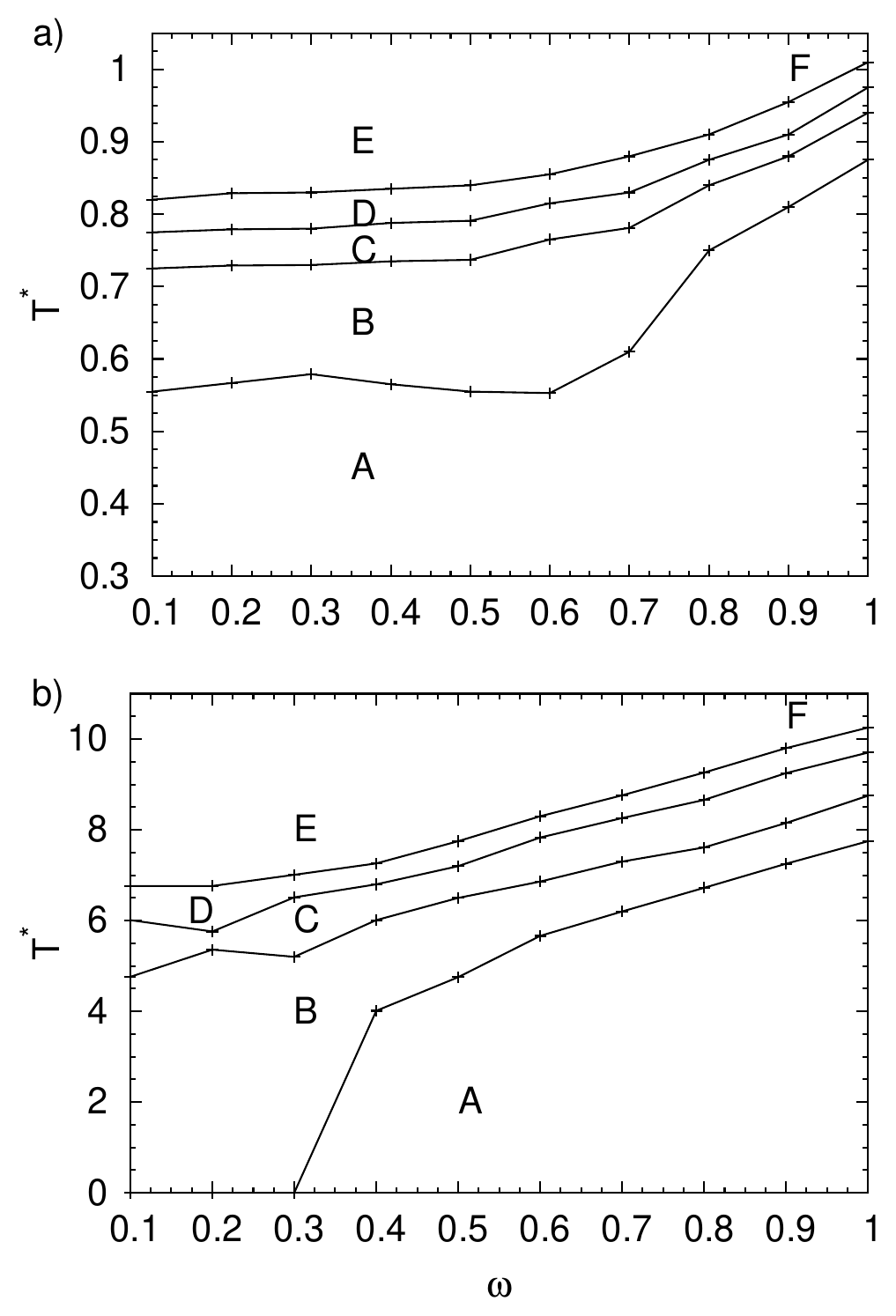}

\protect\caption{Phase diagram in the $(\omega,T^{*})$ parameter space that shows
highest probability of finding different counts of clusters for off-lattice
(from ref. \citet{PhysRevE.84.011806}) model (a) and lattice model
(b). A - 2-cluster, B - 3-cluster, C - 4-clusters, D - 5-clusters,
E - 6-clusters, F - disordered. \label{fig:fig9}}
\end{figure}

The value of R depends on both the chain length and microarchitecture,
but to a good approximation it is equal to about 6.5 \citep{Lewandowski2008}.
In figure (\ref{fig:fig9}) we present both the results for off-lattice
model (a) and the results for the lattice model (b), using the $(10-10)_{12}$
microarchitecture. In both cases most probable structure remains the
same: 2, 3, 4, 5, 6-clusters. Comparing maxima of probability of obtaining
certain clusters we can observe that in both cases those maxima shift
to highers temperatures as the compatibility increases. The most significant
difference is for $\omega<0.4$. In the previous work (ref. \citep{PhysRevE.84.011806})
for lower temperatures and for all values of $\omega$ the most probably
structure was the 2-cluster. In this paper for lower $T^{*}$'s and
$\omega<0.4$ the most probable structure is the 3-cluster. However,
if we compare the transition temperature from the B (2-cluster) region
to the C (3-cluster) region at $\omega=0.1$ then their ratio is about
7 which is close to R \ensuremath{\approx} 6.5 from ref. \citep{Lewandowski2008}.
It is reassuring to notice if we were to rescale the off-lattice phase
diagram by this factor, then both phase diagrams would be very similar
to each other, with the exception of a region which is roughly determined
by the following inequalities: $\omega<0.4$ and $T^{*}<5$.

The reason for this discrepancy may be related to the fact that the
lattice chain is stiffer than the off-lattice chain with the same
number of segments. Indeed, when we compare the off-lattice phase
diagrams for the $(10-10)_{12}$ microarchitecture with that for the
$(6-6)_{20}$ microarchitecture (see ref. \citep{PhysRevE.84.011806}),
we notice that for the $(6-6)_{20}$ microarchitecture the 3-cluster
is the most stable structure at low temperatures. Obviously, the 6-block
of the $(6-6)_{20}$ microarchitecture is stiffer than the 10-block
of the $(10-10)_{12}$ microarchitecture, and therefore we can conclude
that the block stiffness promotes the 3-cluster structure, as also
observed for the lattice $(10-10)_{12}$ microstructure.

\section{Conclusions}

Two different approaches, off-lattice and lattice Monte Carlo simulations
confirm that multiblock copolymer chain in the globular state forms
variety number of different patchy nanostructures. We observed the
expected behavior of phase coil-to-globule transition upon cooling.
We report thermodynamic and structural properties such as energy,
specific heat, mean square radius of gyration. We constructed the
phase diagram in the ($\omega,\:T^{*}$) space which presents the
same results as for the off-lattice simulation in the case of $\omega>0.4$.
For lower compatibility we report that the most probably phase is
3-cluster.

Finally we answer the questions that we posed in the introduction:
\begin{itemize}
\item the nanostructures obtained in the lattice simulation are the same
as those obtained in the off-lattice simulation
\item the phase diagram obtained in the lattice simulation is mostly similar
to that obtained in the off-lattice simulation, but qualitatively
different for $\omega$\textquoteright s smaller than 0.4 and $T^{*}$\textquoteright s
smaller than 5 (3-cluster is more probable than the 2-cluster). This
discrepancy, however, can be tentatively attributed to a considerable
stiffness of lattice chains in comparison to that of the off-lattice
chains.
\end{itemize}

\section*{ACKNOWLEDGMENTS}

A significant part of the simulations was performed at the Poznan
Computer and Networking Center (PCSS).

\pagebreak{}

\bibliographystyle{plainnat}
\bibliography{lterature}

\end{document}